\documentclass{article}
\usepackage{amssymb,xypic,latexsym}
\usepackage{geometry}
\usepackage{graphicx}

\title{Geometric Aspects of Multiagent Systems}
\author{Timothy Porter \\ School of Informatics,\\ University of Wales
  Bangor,\\ Bangor,\\ Gwynedd, LL57 1UT, \\Wales, U.K.\\
Email: tporter@bangor.ac.uk} 

\newtheorem{proposition}{Proposition}

\newtheorem{lemma}{Lemma}

\begin{document}

\maketitle

\begin{abstract} Recent advances in Multiagent Systems  (MAS) and
Epistemic Logic within Distributed Systems Theory, have used
various combinatorial structures that model both the geometry of
the systems and the Kripke model structure of models for the
logic. Examining one of the simpler versions of these models,
interpreted systems, and the related Kripke semantics of the logic
$S5_n$ (an epistemic logic with $n$-agents), the similarities with
the geometric / homotopy theoretic structure
 of groupoid atlases is striking.  These latter objects arise in problems within algebraic K-theory, an area of algebra
 linked to the study of decomposition and normal form theorems in linear algebra.  They have a natural well structured
 notion of path and constructions of path objects, etc., that yield a rich homotopy theory.

In this paper, we examine what an geometric analysis of the model
may tell us of the MAS. Also the analogous notion of path will be
analysed for interpreted systems and $S5_n$-Kripke models, and is compared to
the notion of `run' as used with MASs. Further
progress may need adaptions to handle $S4_n$ rather than $S5_n$
and to use directed homotopy rather than standard `reversible'
homotopy.
\end{abstract}
\pagebreak
\begin{center}
\Large\textbf{Geometric Aspects of Multiagent Systems\\
\vspace{5mm}
Timothy Porter }

\end{center}
\section{ Introduction}
In many studies of distributed systems, a multiagent model is
used.  An agent is a processor, sensor or finite state machine,
interconnected by a communication network with other `agents'.
Typically each agent has a local state that is a function of its
initial state, the messages received from other agents,
observations of the external environment and possible internal
actions.  It has become customary when using formal models of
distributed systems to use modal epistemic logics as one of the
tools for studying the knowledge of such systems. The basic such
logic for handling a system with $n$-agents is one known as
$S5_n$. Unless the system is very simple the actual logic will be
an extension of that basic one, that is, it may have more axioms.
For instance, the way the various agents are connected influences
the logic in subtle ways. Suppose that agent 1 sends all its
information immediately to agents 2 and 3, then if we denote by
$K_i\phi$, the statement that agent $i$ `knows' proposition
$\phi$, we clearly expect within the logic of that system that
$K_1\phi\Rightarrow K_2\phi \wedge K_3\phi$.

The logic $S5_n$ is obtained from ordinary propositional logic by
adding `knowledge operators', $K_i$ as above. (In the literature
the notation $K_i\phi$ is often replaced by $\Box_i\phi$.)  It
models a community of \underline{ideal} knowledge agents who have
the properties of veridical knowledge (everything they know is
true), positive introspection (they know what they know) and
negative introspection (they know what they do not know). These
properties are reflected in the axiom system for the logic.  The
axioms include all propositional tautologies, plus the schemata of
axioms: $K_i(\phi\Rightarrow \psi) \Rightarrow (K_i\phi\Rightarrow
K_i \psi)$, $K_i\phi \Rightarrow \phi$, $K_i\phi \Rightarrow
K_iK_i\phi$, and $\neg K_i\neg \phi \Rightarrow K_i\neg K_i\neg
\phi$, where $i \in A$, the set of `agents'.  (We will see an
alternative presentation of the logic later on.) Two comments
worth making are (i) several of these axioms and in particular the
last one - negative introspection - are considered computationally
infeasible and (ii) ideas such as common knowledge (represented by
an additional operator, $C$) can be introduced to give a richer
extended language.  Here however we will be restrict attention
largely to models for $S5_n$ and extensions that may reflect the
geometry of the system being modelled. How is this `epistemic
analysis´ used in practice?  We mention three examples. One is given in
\cite{MvdH} (\S 1.9) as due to Halpern and Zuck.  It shows the way in which
epistemic operators give compact and exact specifications of protocols that
are verifiably `safe'. Another worth mentioning is the analysis of AI data /
knowledge searches, such as the Muddy Children problem (cf.  Lomuscio and
Ryan, \cite{ALMR4}).  Finally the study of knowledge based programming,
\cite{HM:UCKBP99}, in which languages one may require statements such as : if
$i$ knows $\phi$, set $x = 0$, by formalising what `knows' means in this
context requires analyses of this type.  The book, \cite{MvdH}, and several of the papers cited here contains numerous further examples.

The classical models for multimodal logics, and for $S5_n$ and its extensions in particular, are combinatorial models known as Kripke frames and, for $S5_n$, Kripke equivalence frames.  These consist of a set $W$, called the \emph{set of possible worlds}, and $n$-equivalence relations $\sim_i$, one for each agent.  The interpretation of $\sim_i$ is that if $w_1$, $w_2$ are two possible worlds and $w_1\sim_i w_2$, then agent $i$ cannot tell these two worlds apart.
In a series of papers and books (see in particular \cite{FHMV95}) Fagin, Halpern, Moses and Vardi, in various combinations, have put forward a simpler combinatorial model known as an \emph{interpreted system}.  These have the same formal expressive power as Kripke frames, but are nearer the intuition of interacting agents than is the more abstract Kripke model.

In each case the underlying frame / set of global states, has a very similar
combinatorial structure to that underlying a structure, \emph{global actions}
(or \emph{groupoid atlases}), introduced by A. Bak, \cite{AB1,AB2}.  These
arose from an analysis of algebraic problems related to the solution of
systems of linear equations over arbitrary rings. (The mathematical area is
\emph{algebraic K-theory} and lies at the interface between algebraic topology
and algebra / algebraic geometry).  Any action of a group on a set leads to a
set of orbits.  These are the equivalence classes for an equivalence by a
`reachability' or `accessibility' relation generated by the group
action. (Translating and weakening to a monoid action, one has a variant of
the reachability of states in a finite automaton.)  In a global action, the
set $X$ is divided up into a family of patches, each of which has a  group
attached, which acts on that patch (see below for the more detailed
definition).  If the patches all coincide the resulting `single domain global
action' is essentially a set with a collection of (possibly independent/
possibly interacting) group actions.  As group actions yield groupoids by a
well known construction, and the resulting equivalence relations are also
groupoids, a useful generalisation of global actions is that of groupoid atlas
introduced by Bak,  Brown, Minian and the author, \cite{GAGA}. These therefore
present a context in which both the algebraic ideas and the logical models of
$S5_n$ can fit.  Moreover both global actions and groupoid atlases have a rich
homotopy theory.  This homotopy theory is based on a notion of path that,
suitably modified, bears an uncanny resemblance to that of the `runs'
considered in multiagent systems, but seems to be better structured and, in
fact, more computationally realistic.

The point of this paper is to examine these models in some more detail and to
start the analysis of the necessary modifications to the global
action/groupoid atlas homotopy theory that will allow its application to the
problem of the \underline{geometric} analysis of multiagent systems: how does
the geometry of a multiagent system influence its inherent logic and thus its
computational ability?  The author's hope is that such an analysis will aid in
three specific problem areas: firstly, any analysis of systems such as these
hits the combinatorial explosion problem, the effective state space is too
large for efficient search to be implemented.  By reducing the search space
via homotopical methods, it is expected that some progress in this can be
achieved.  Next, some distributed systems can be modularised thereby aiding
verification that their description and behaviour matches their specification.
This modularisable attribute should be identifiable by a combination of
algebraic and geometric tools.  A related question here is as to whether or
not it is better to group a set of agents together as one `super-agent' and
under what condition can this be done without changing the behaviour of the
system for the worse.\footnote{Our models of agents tend to work  as if they
  are given `atomic' entities, however if they interact or if they themselves consist of `subagents', (processors), a different grouping into `full agents' may be beneficial to analysis, optimisation and verification.}  This, of course, presupposes a mechanism for  comparison of MASs with different numbers of agents, a point to which we will return.  The final hope is that a geometric homotopical overview may aid in the description and handling of \emph{knowledge evolution} without MAS.

\medskip

\textbf{Acknowledgements}

The research in this paper would not have been possible without conversations
with Tony Bak, George Janelidze, and various participants at the 2000 category
theory conference at the Villa Olmo, Como, and therefore without the assistance of ARC grant Number 859 and
INTAS grants (INTAS 93-436 and 97-31961). It has also benefitted from
conversations and discussions with Eric Goubault, Jean Goubault-Larrecq and
other members of the Geometric Methods in Theoretical
Computer Science group.

\section{Preliminaries.}

(As references for basic modal logic, try Kracht, \cite{kracht}, Meyer and van
der Hoek, \cite{MvdH} and Blackburn, de Ryke and Venema, \cite{BdRV}.)  In the
following, at least to start with, there will be $n$-agents and $A = \{1,2, \ldots, n\}$ will denote the set of such `agents'.

\textbf{$S5$ and $S5_n$.}

To introduce these logics fairly formally we suppose given a set of variables and form a language $\mathcal{L}_\omega(n)$ given by
$$\phi ::= p_\lambda ~|~ \bot ~|~ \neg \phi ~|~ \phi_1 \vee \phi_2 ~|~ M_i\phi$$
where the $p_\lambda$ are the propositional variable ordered by the finite ordinals $\lambda$, and $M_i$ is a modality for each agent $i=1, \ldots , n$.

\medskip

In contrast to some treatments, we are using operators, $M_i$,  corresponding to ``possibility'' , rather than  ``knowledge'' operators, i.e. we interpret $M_i\phi$  as ``agent $i$ considers $\phi$ is possible''.  The relation with $K_i\phi$ (``agent $i$ knows $\phi$'') is $M_i = \neg K_i\neg \phi$, ``agent $i$ does not know that $\phi$ is false''.  For computational purposes these may be expected to yield different methods, since $\neg$ is not well behaved computationally, however for this paper we will not be considering computational/ implementational problems, so the $M$ v. $K$ debate need not concern us greatly.

A logic in $\mathcal{L}_\omega(n)$ is any set $\Lambda$ of $\mathcal{L}_\omega(n)$-formulae such that
\begin{itemize}\item $\Lambda$ includes all $\mathcal{L}_\omega(n)$-formulae that are instances of tautologies, \\
\hspace*{-.8cm}and
\item $\Lambda$ is closed under the inference rule \\
\centerline{if $\phi$, $\phi\to \psi \in \Lambda$ then $\psi \in \Lambda$\hspace*{1cm}}
i.e. detachment or \emph{modus ponens}
 \end{itemize}

 The logic is \emph{uniform} if it is closed under the rule of uniform substitution of $\mathcal{L}_\omega(n)$-formulae for propositional variables and is \emph{normal} if it contains the schemata\\
$(K) \quad M_i(\psi \vee \chi) \to M_i(\psi)\vee M_i(\chi)$
\\
$(N)\quad \neg M_i(\bot)$\\
and monotonicity (for each $i$):\\
\centerline{if $\psi \to \chi \in \Lambda$ then $M_i \psi \to M_i \chi \in \Lambda$.}\\
 As is well known, $S5_n$ is defined to be the smallest normal logic in $\mathcal{L}_\omega(n)$ containing \\
$(T)_i \quad \phi \to M_i\phi$\\
$(4)_i \quad M_iM_i \phi \to M_i \phi$\\and\\
$ (B)_i\quad \phi \to K_i M_i\phi$,\\
(so is $K4.BT$ or $S4.B$  in the notation used in Kracht, \cite{kracht}, p.72).

\medskip
Of course $\phi\to\psi$ is shorthand for $\neg\phi\vee \psi$

As we are in the classical rather than the intuitionistic case, it is easy to rewrite this in terms of $K_i$ instead of $M_i$.  The logic $S5_1$ is usually called $S5$.

The related logic $S4_n$, mentioned earlier,  does not require the schemata $(B)_i$.

The usual semantics of $S5_n$ is given by Kripke equivalence frames and models .

\medskip

\textbf{Kripke equivalence frames}

An \emph{equivalence frame}  (or simply \emph{frame}) $F = (W, \sim_1, \ldots,
\sim_n)$ consists of a set $W$ with, for each $i \in A$, an equivalence
relation $\sim_i$ on $W$.  Elements of $W$ are called \emph{worlds} and are
denoted $w$, $w^{\prime}$, etc.  We will write $[w]_i$ for the equivalence class of
the element $w \in W$ for the $i$\textsuperscript{th} equivalence relation,
$\sim_i$.  A Kripke frame is very like a labelled transition system, but it
has equivalence relations rather than partial orders as its basic relational structure.
  The logic gives a semi-static view of the system.  To get a dynamic aspect one needs to look at knowledge evolution, cf. for example, Lomuscio and Ryan, \cite{ALMR4}.

An \emph{equivalence (Kripke) model} (or simply \emph{model}) $M = (F, \pi)$ is a frame $F$ together with  a relation
$$R_\pi\subseteq P\times W,$$ where $P = \{p_\lambda ~:~ \lambda \in \mathbb{N}\}$.  This relation yields an \emph{interpretation}
$$\pi_R  : W \rightarrow \mathcal{P}(P),$$
which interprets as : $\pi_R(w)$ is the set of basic propositions ``true'' at $w$, or dually a \emph{valuation}
$$_L\pi : P \to \mathcal{P}(W)$$giving : $_L\pi(p)$ is the set of  worlds at which $p$ is
``true''. Of course $\pi_R$ and $_L\pi$ contain the same information and will be merged in notation to $\pi$ when no confusion  will result.

A  weak map or weak morphism of frames $f : F \rightarrow F^{\prime} = (W^\prime,
 \sim_1^\prime, \ldots,\sim_n^\prime)$ is a function $f : W \rightarrow
 W^\prime$ such that for each $i$,
\begin{center} if $w\sim_iw^\prime$, then $f(w)\sim_i^\prime
  f(w^\prime).$\end{center}
 The map $f$  will give a map of models $f :(F,\pi) \rightarrow
(F^\prime, \pi^\prime)$ if
$$\xymatrix{W\ar[rr]^f\ar[dr]_\pi&&W^\prime\ar[dl]^{\pi^\prime}\\
& \mathcal{P}(P)&}$$
commutes.

\medskip

Weak maps are too weak to react well with the logic so a stronger notion of bounded morphism (or p-morphism) is also used.

A weak morphism $f : F \to F^\prime$ of frames is \emph{bounded} if for each $i$, $1 \leq i \leq n$ and $u \in W$, $v^\prime \in W^\prime$,
$$f(u) \sim_i^\prime v^\prime \mbox{ if and only if there is a }v \in W \mbox{ with } f(v) = v^\prime \mbox{ and } u \sim_i v.$$

\medskip

\textbf{Remark:}

 A discussion of some of the properties of the resulting categories of frames
 and weak maps (or of frames and bounded maps) can be found in \cite{TP1}.
  In this paper we will not be considering bounded morphisms nor models in any
 great detail due to restrictions on space.

\medskip

\textbf{Global states for interpreted systems}

Interpreted systems were first proposed by Fagin, Halpern, Moses and Vardi, \cite{FHMV95} to model distributed systems.  They give simple combinatorial models for some of the formal properties of multiagent systems.  As before one has a set, $A = \{1,2, \ldots, n\}$, of agents, and now one assumes each agent $i$ can be in any state of a set $L_i$ of local states.  In addition one assumes given a set $L_e$ of possible states of the `environment'.  More formally:
 
A \emph{set of global states} (SGS) for an interpreted system is a subset $S$ of the product $L_e \times L_1 \times \ldots \times L_n$ with each $L_e$, $L_i$ non-empty.  If $S = L_e \times L_1 \times \ldots \times L_n$, then the SGS is called a \emph{hypercube}, cf.  \cite{ALMR1}.

The idea behind allowing the possibility of considering a subset and not just the whole product is that
some points in $\prod \underline{L} = L_e \times \prod_{i = 1}^{n}  L_i$ may
not be `feasible', because of explicit or implicit constraints present in the
multiagent system (MAS).  The explicit way these constraints might arise is
usually not considered central for the general considerations of the
multimodal logic approach to MASs, yet it seems clear that it represents the
interconnection of the network of agents and, if the local states are the
states of a finite state automaton, questions of reachability may also  arise.
This will be where the `topology' of the MAS is most clearly influencing the
combinatorial topology of the model.  As a simple example, suppose we have
agent 1 acts solely as a sensor for agent 2, so anything agent 1 knows, agent
2 automatically knows, $K_1\phi     \Rightarrow K_2\phi$.  The effect of this
can be illustrated where $L_1$ has two local states, $x_1$ and $x_2$.  In
$x_1$, $p$ is true; in $x_2$, $\neg p$ is true.  Suppose $L_2$ has 5 local
states, $y_1, \ldots, y_5$, and $p$ is true only in $y_1$ and $y_2$, $\neg p$
being true in the remainder.   Then $S = 
\{(x_1,y_1),(x_1,y_2),(x_2,y_3),(x_2,y_4),(x_2,y_5)\} $ is as large a SGS (or
more precisely interpreted system as the valuation plays a role) as one can
get within this setting.  The situation mentioned earlier, $K_1\phi \Rightarrow K_2\phi\wedge K_3\phi$, will lead to a similar 3-dimensional example.  The link between the structure of the SGS, the logic inherent in the interrelations between agents and the computational power of the system is subtle, see \cite{ALMR3} for a set of examples.  Other restrictions may also play a role.  Agents may share resources, e.g. in a context where they need to access a distributed database and one agent may block another from an otherwise feasible transition.\footnote{For simplicity, it is assumed that each local agent is a reversible transition system. Thus if a transition $s \to s^\prime$ can occur in $L_i$, then  $s^\prime \to^\ast s$ as well, i.e. we can get back from $s^\prime$ to $s$ by some sequence of transitions, even if this requires reinitialising $L_i$, but any given transition in $L_i$ may not be feasible at some state of $S$, being blocked by the actions of other agents, whence the complication of the  system.}

Any SGS yields a Kripke frame.  If we write $\underline{L} = (L_e, L_1,
\ldots, L_n)$ and $(S, \underline{L})$ for an SGS based on  $\underline{L}$,
then set $\mathcal{F}(S, \underline{L})$ to be the frame with $S$ as its set
of possible worlds with $\sim_i$ defined by: $$\underline{\ell}\sim_i
\underline{\ell}^\prime\quad\Leftrightarrow \quad\ell_i = \ell^\prime_i,$$ i.e. $\underline{\ell}$ and
$\underline{\ell}^\prime$ correspond to the same local state for agent $i$. For
simplicity we will assume that $L_e$ is a singleton set.

There are notions of weak map and bounded map of SGSs and an adjoint equivalence between the categories of frames and those of SGSs modulo a notion of essential equivalence.  If $F =  (W, \sim_1, \ldots, \sim_n)$ is an equivalence frame, then for each agent $i$, let $W_i = W /\sim_i$, be the set of equivalence classes of elements of $W$ for the relation, $\sim_i$, and set $\underline{W} = (W_1, \ldots, W_n)$.  There is a `diagonal' function
$$\Delta : W \to \prod\underline{W},$$
given by $$\Delta(w) = ([w]_1, \ldots, [w]_n)$$
and $(\Delta W, \underline{W})$ is an SGS.  Setting $\mathcal{G}(F) = (\Delta W, \underline{W})$ gives the functor, left adjoint to $\mathcal{F}$, that is used in \cite{TP1} to prove the equivalence mentioned above.

\medskip

\textbf{Mathematical Interlude:  Global Actions and Groupoid Atlases.}

A very similar structure to a Kripke equivalence frame is that of a global action, see \cite{AB1,AB2}.  Their generalisation in \cite{GAGA} to groupoid atlases gives a context where both Kripke equivalence frames and global actions coexist and it is a situation with a well defined and quite well behaved homotopy theory, therefore it yields a potential tool for the geometric analysis of MASs.

The prime example of a global action is a set $X$ with a family of groups acting on it.  In particular if $G$ is a group (in the usual mathematical sense) then given a family of subgroups $\{H_i : i \in I\}$ of $G$, we can consider the actions of each $H_i$ on the set of elements of $G$ by left multiplication.  The important point to note is that the different subgroups $H_i$ may be related, e.g. we may have $H_i \subseteq H_j$, which implies structural relationships between the equivalence relations generated by the actions of $H_i$ and $H_j$.  In more detail, we have for each $i \in I$, an equivalence relation $\sim_i$ on the set of elements of $G$ defined by
$$x\sim_i y \mbox{ if and only if there is some }h_i\in H_i \mbox{ with } x = h_iy.$$
If $H_i\subseteq H_j$, then $x\sim_i y$ implies $x\sim_j y$, which is exactly the sort of relationship that results from `knowledge passing' within a MAS, (cf. \cite{ALMR2}).  In a global action or groupoid atlas, this relationship is explicitly specified from the start.  The example above is a single domain global action as there is one set on which all the groups act.  The general form assumes only that the groups act on subsets of the `domain'.  This adds additional flexibility and adaptability to the concept.  (In addition to the notes, \cite{GAGA}, the original definition and discussion of global actions can be found in \cite{AB1,AB2}.)

A \emph{global action} $\mathfrak{A}$ consists of a set $X_\mathfrak{A}$, together with a family $\{(G_\mathfrak{A})_\alpha\curvearrowright  (X_\mathfrak{A})_\alpha ~|~ \alpha \in \Phi_\mathfrak{A}\}$ of group actions on subsets $(X_\mathfrak{A})\alpha \subseteq X_\mathfrak{A}$.  The various \emph{local groups} $(G_\mathfrak{A})_\alpha$ and the corresponding subsets (\emph{local patches}), $(X_\mathfrak{A})_\alpha $, are indexed by the index set, $\Phi_\mathfrak{A}$, called the \emph{coordinate system} of $\mathfrak{A}$.  This set $\Phi_\mathfrak{A}$ is equipped with a reflexive relation, written $\leq$, and it is required that

-  if $\alpha \leq \beta$ in $\Phi_\mathfrak{A}$, then $(G_\mathfrak{A})_\alpha$ leaves $(X_\mathfrak{A})_\alpha \cap(X_\mathfrak{A})_\beta $ invariant (so  $(X_\mathfrak{A})_\alpha \cap(X_\mathfrak{A})_\beta $ is a union of equivalence classes for the $(G_\mathfrak{A})_\alpha $-action), and\\
-  there is given for each pair $\alpha \leq \beta$, a group homomorphism
$$(G_\mathfrak{A})_{\alpha \leq \beta} : (G_\mathfrak{A})_\alpha \to (G_\mathfrak{A})_\beta$$
such that if $\sigma \in (G_\mathfrak{A})_\alpha$ and $x \in (X_\mathfrak{A})_\alpha \cap(X_\mathfrak{A})_\beta $, then
$$\sigma x = (G_\mathfrak{A})_{\alpha \leq \beta}(\sigma)x.$$

This second axiom says that if $\alpha$ and $\beta$ are \emph{explicitly}
related and their domains intersect then the two actions are related on that
intersection. Again this is the sort of structural compatibility that arises in MASs, except that, as so far considered, interpreted systems, etc. do not allow for the `multi-patch' setting.

Any global action yields on each $(X_\mathfrak{A})_\alpha$ an equivalence relation due to the $(G_\mathfrak{A})_\alpha$-action.  The equivalence classes (local orbits or local components) for these `local equivalence relations' form a structure that is sometimes useful, regardless of what group actions are used, i.e. we need the local equivalence relations rather than the local groups that were used to derive them.  As both equivalence relations and group actions yield groupoids (small categories in which all the morphisms are isomorphisms), it is convenient to adapt the notion of global actions to give a generalisation which handles the local equivalence relations as well.  This generalisation is called a groupoid atlas in \cite{GAGA}. Fuller details of that transition are given in that source.

\medskip

A \emph{groupoid atlas} $\mathfrak{A}$ on a set $X_\mathfrak{A}$ consists of a family of groupoids $(G_\mathfrak{A})_\alpha$ defined with object sets $(X_\mathfrak{A})_\alpha$ which are subsets of $X_\mathfrak{A}$.  These \emph{local groupoids} are indexed by an index set $\Phi_\mathfrak{A}$, called the \emph{coordinate system} of $\mathfrak{A}$, which is equipped with a reflexive relation, written $\leq$.  This data is required to satisfy:\\
(i) if $\alpha \leq \beta$ in $\Phi_\mathfrak{A}$, then $(X_\mathfrak{A})_\alpha \cap (X_\mathfrak{A})_\beta$ is a union of components of $(G_\mathfrak{A})_\alpha$, i.e. if $x \in(X_\mathfrak{A})_\alpha \cap (X_\mathfrak{A})_\beta$, and $g \in (G_\mathfrak{A})_\alpha$, $g : x \to y$, then $y \in (X_\mathfrak{A})_\alpha \cap (X_\mathfrak{A})_\beta$ ;\\
and\\
(ii) if $\alpha \leq \beta$ in $\Phi_\mathfrak{A}$, then there is given a groupoid morphism,
$$(G_\mathfrak{A})_\alpha\Big|_{(X_\mathfrak{A})_\alpha \cap (X_\mathfrak{A})_\beta} \to (G_\mathfrak{A})_\beta\Big|_{(X_\mathfrak{A})_\alpha \cap (X_\mathfrak{A})_\beta},$$
defined between the restrictions of the local groupoids to the intersections, and  which is the identity on objects.

\medskip

\textbf{Example 1. (From Kripke frames to Groupoid Atlases.)}

Let $X$ be a set and $\sim_i$, $i = 1,2, \ldots, n,$ $n$ equivalence relations on $X$. Then $F = (X, \sim_1, \ldots, \sim_n)$ is a Kripke frame, but also, if we specify the local groupoids
$$G_i:= \mbox{ Objects } X, \mbox { arrows } x_1 \to_ix_2 \mbox { if and only if } x_1\sim_i x_2,$$
 and $\Phi$ to be discrete, i.e. ``$~\leq ~$" = `` = '', we have a simple
 groupoid atlas, $\mathfrak{A} (F)$, (cf. example 2 \S2 of \cite{GAGA}).  In fact later we will introduce a second method for turning a frame into a groupoid atlas.

\medskip

\textbf{Example 2. The Line}

The simplest non-trivial groupoid is $\mathcal{I}$.  This is the groupoid corresponding to the Kripke frame $W = \{0, 1\}$,
$\sim$ =  the  indiscrete / universal equivalence relation so $1\sim 0$.  (If the
number of equivalence relations / agents is needed to be kept constant, then
set $\sim_i ~=~ \sim$ for $i = 1, \ldots ,n$.  This is sometimes useful, but
should not concern us too much for the moment.

The line, $\mathfrak{L}$, is obtained by placing infinitely many copies of
$\mathcal{I}$ end to end, so \\
\begin{eqnarray*}
|\mathfrak{L}|&:=& \mbox{the set}, \mathbb{Z},  \mbox{of integers}\\
\Phi & := &\mathbb{Z}\cup \{-\infty \}, \mbox{ where } - \infty \leq -\infty, ~ -\infty <
  n \mbox{ for all } n \in \mathbb{Z} \mbox{ and } n \leq n,
\end{eqnarray*}
but that gives all related pairs.

\medskip

\textbf{What about models?}

The above construction (Example 1 and later on its variant) gives us a way to think of Kripke frames and SGSs as groupoid atlases, but they do not directly consider the interpretations / valuations that are needed if Kripke models and interpreted systems are to be studied via that combinatorial gadgetry.  Given a frame $F = (W, \sim_1, \ldots,
\sim_n)$  and an interpretation $$\pi : W \to  \mathcal{P}(P),$$we can get a bounded morphism of frames
$$\pi_\ast : F \to S^\Lambda_P$$
for $\Lambda = S5_n$ or an extension.  This frame $S^\Lambda_P$ is the \emph{canonical frame} for the logic $\Lambda$ with the given set of variables $P$.  Its `possible worlds' are the $\Lambda$-maximal sets of  $\mathcal{L}_\omega(n)$-formulae, and $\pi_\ast$ assigns to a world $w$ the set of $\phi$ such that $(F,w)\models_\pi \phi$, i.e. the set of $\phi$ valid at the world $w$ given $\pi$ as interpretation, (cf. \cite{kracht} p.63).  Thus, if the homotopy theory of frames / SGSs informs us about their `geometry', the homotopy theory of frames \emph{over} $S^\Lambda_P$ should inform us of the corresponding `geometry' of the $\Lambda$-models in each context.  Because of this and our relative ignorance of `homotopy over' in \emph{this context}, we will put models aside for this paper and concentrate on frames and SGSs.

\section{Morphisms, Runs, Curves and Paths.}

In the previous section, we have seen that groupoid atlases form a class of structures that encompasses Kripke equivalence frames as well as more general objects such as global actions.  This by itself need not be useful.  The general notion of a categorical model of a situation demands that serious attention be paid to the morphisms. We have seen that Kripke frames and interpreted systems have both weak morphisms and bounded morphisms available for use, the latter preserving more of the internal logic.  We therefore need to consider morphisms of groupoid atlases.  The payoff will be if the known function space structure on certain classes of morphisms between groupoid atlases (cf. \cite{AB1,AB2}) can allow a similar structure to be made available for models of MASs.

\medskip

A function $f : |\mathfrak{A}|\to |\mathfrak{B}|$ between the underlying object sets of two groupoid atlases is said to support the structure of a weak morphism if it preserves \emph{local frames}, (the term comes from the original work on global actions and is not connected with the model theoretic meaning). Here a local frame in $\mathfrak{A}$ is a set $\{x_0, \ldots, x_p\}$ of objects in some connected component of some $(G_\mathfrak{A})_\alpha$,  i.e. $\alpha \in \Phi$ and there are arrows $g_i :x_0\to x_i$ in $(G_\mathfrak{A})_\alpha$ for $i = 1, 2, \ldots, p$.   The function $f$ preserves local frames if for $\{x_0, \ldots, x_p\}$ is a local frame in $\mathfrak{A}$ then $\{f(x_0), \ldots, f(x_p)\}$ is a local frame in $\mathfrak{B}$.

Any weak morphism of Kripke frames will give a weak morphism of the corresponding groupoid atlases, but not conversely since if $\mathfrak{A} = (W, \sim_1, \ldots,
\sim_n)$ and $\mathfrak{B} = (W^\prime, \sim^\prime_1, \ldots, \sim^\prime_n)$, the notion of weak morphism of groupoid atlases allows $f : W \to W^\prime$ to ignore which agents are involved, i.e.
$\{w_0, \ldots , w_p\}$ is a local frame in $\mathfrak{A}$ if there is some
agent $i$ such that $w_0 \sim_iw_k$, $k = 1, \ldots, p$, so agent $i$
considers these worlds equivalent; if $f$ preserves this local frame, then
there is some agent $j$ such that $f(w_0), \ldots , f(w_p)$ are considered
equivalent by that agent.  Note however that agent $j$ need not be the same
agent as agent $i$, nor necessarily have the same position in the lists of
agents if the sets of agents in the two cases are represented by disjoint
lists. In fact the number of agents in the context of the two  Kripke frames
did not actually need to be the same for a weak morphism to exist between
them.  This added flexibility would seem to be essential when discussing
modularisation, as mentioned above, but also for interacting MASs and the
resulting interaction between the corresponding epistemic logics, however note
these are \emph{weak} morphisms so the link with the logic here is fairly
weak.  There is a notion of strong morphism of groupoid atlases, but this will
not be strong enough either for the logic.  In fact Bak's notion of stong morphisms in this particular context reduces to that of weak morphisms.  The difference is that in a weak morphism the `reasons', i.e. the elements $g_i$ above,  that a set of objects forms a local frame is not considered a part of the data of the morphism, with strong morphisms this data \emph{is} recorded.  In the groupoid atlases derived from Kripke frames, there is only one `reason' possible.  If it exists, it is unique! Thus the difference between the two types of morphism can be safely ignored for the moment. We hope to return to morphisms of groupoid atlases that correspond to bounded morphisms of Kripke frames and SGSs in a future paper.

\medskip

\textbf{Runs }(cf. \cite{MvdH}, p. 59).

A \emph{run} in a Kripke model $M = (F, \pi)$ associated with a distributed system / MAS is simply a finite or infinite sequence of states $s^{(1)}, s^{(2)}, \ldots, $ with $s^{(i)} \in S$, the set of possible worlds of $F$, here being thought of as being an SGS for an interpreted $(S, \underline{L})$.

A weakness in this definition is that no apparent restriction is put on adjacent states in a run. This thus ignores essential structure in the SGS,  and any link between runs and  morphisms is not immediately clear. Because of this, we will take the view that as formulated, this notion of `run' is not quite adequate for the analysis of these systems. It needs refining, bringing it nearer to the mathematical notion, not just for {\ae}sthetic reasons but also because it does not do the job for which it was `designed'!  It does work well in some situations however. If we, for the moment, write $x\to x^\prime$ to mean $x\sim_i x^\prime$ for some $i$ and then extend to the corresponding category (reflexive, transitive closure) to give $x\to ^* y$, then (cf. again  \cite{MvdH}, p. 60), for hypercubes with more than one agent, any two states are related via $\to^*$. 

\begin{lemma}~~\\
If $M$ is a hypercube SGS associated to a distributed system with more than one agent, then given any states $s, t$ in $S$, $s \to^*t$.
\end{lemma}

\textbf{Proof}

If $\underline{s} = (s_1, \ldots, s_n)$ and $t = (t_1, \ldots , t_n)$ then
$$(s_1, \ldots, s_n)\to (t_1,s_2, \ldots, s_n) \to (t_1, \ldots , t_n).$$
The first arrow comes from $\sim_i$, and $i \neq 1$, the second from
$\sim_1$. \hfill $\blacksquare$
\medskip

Of course, this argument may fail if $(S, \underline{L})$ is not a hypercube
as simple examples show.

\begin{proposition}~~\\
If $M$ is a Kripke frame or SGS, considered, as above, as a groupoid atlas, then
any weak morphism $$f : L \to M$$ for which $f(n) = f(0)$ for all $n < 0$,
determines a run $s^{(i)} = f(i)$ in $S$  \hfill $\blacksquare$
\end{proposition}

Often the form of the set of global states is not specified that precisely.
Sometimes local transition functions are used so that the $L_i$ are
transformed into ``local transition systems'' with the actions involved being
coupled with each other (cf. for instance, the $\mathcal{VSK}$ systems of
Wooldridge and Lomuscio \cite{AL&MW}).  The feasible runs would seem in any
case to be those for which  $s^{(i+1)}$ is reachable from $s^{(i)}$
 so there is some set of transitions in the various agents that leads from
$s^{(i)}$ to $s^{(i+1)}$, or precisely:

\centerline{ a run $(s^{(k)})$ is \emph{feasible} if for each $k = 1,2,
  \ldots,$ $s^{(k)}\to^* s^{(k+1)}.$}

Of course, hidden within this notion is a certain potential for concurrency.  We do not specify in $(s^{(k)})$ how $s^{(k)}$
becomes $s^{(k+1)}$ except that by some set of local transitions within the
state spaces of the different agents, components of $s^{(k)}$ have changed to
become those of $s^{(k+1)}$  and at all times \emph{the resulting intermediate
  list of local states is a valid one}, i.e. is a list of global states within $S$.

Looking at finer granularity, assume that $s^{(k)}$  and $s^{(k+1)}$
 are not linked directly, i.e. $s^{(k)}\to^* s^{(k+1)}$ but it is not the case
 that $s^{(k)}\to s^{(k+1)}$.

If $\underline{s}$, $\underline{t}$ are two states in $(S, \underline{L})$, we
will write
$$HC(\underline{s},\underline{t}) = \{ \underline{x}\in \prod \underline{L}:
\mbox{ for each }i, 1 \leq i \leq n, x_i = s_i \mbox{ or } t_i\}$$
and say this is the \emph{hypercube interval} between $\underline{s}$ and
$\underline{t}$ .

\begin{proposition}~\\
Suppose $(s^{(k)})$ is a run in $\mathfrak{M} = (S, \underline{L})$. If $HC(s^{(k)},s^{(k+1)}) \subset S$ for each $k$, then there is a morphism
$$f : \mathfrak{ L} \to \mathfrak{M}$$
of groupoid atlases satisfying $f(n) = s^{(1)}$, $ n\leq 2$
$$f(2k) = s^{(k)}, \quad k = 1,2, \ldots $$
\hfill$\blacksquare$
\end{proposition}
In other words, if at each stage, the hypercube interval between adjacent states of a run is contained in $S$, we can replace $(s^{(k)})$ by a curve.  Within each hypercube interval, there are many possible concurrent paths between adjacent states of the run.  We therefore have not only that a `curve' can be given to represent the run, but the different representing curves are in some sense `homotopic', i.e. equivalent via deformations (or interleaving equivalence).  Of course, the condition is far from being necessary.  If each $s^{(k)}\to^\ast s^{(k+1)}$, we could find a curve, but would not be able to specify it as closely.  The intermediate `odd' points of the curve can be given, up to interleaving equivalence, as in the earlier lemma.

The precise definition of a curve in a groupoid atlas is given as follows:

If $\mathfrak{A}$ is a groupoid atlas, a \emph{(strong) curve} in $\mathfrak{A}$ is a (strong) morphism of groupoid atlases
$$f : \mathfrak{L} \to \mathfrak{A},$$
so for each $n$, one gets a $\beta \in \Phi_\mathfrak{A}$ and $f(i_m):f(n) \to f(n+1)$ in $(\mathcal{G}_\mathfrak{A})_\beta$, where we have written $f(i_m)$ for $f_\mathcal{G}(i_m)$, where $i_m : m \to m+1$ in $(X_\mathfrak{L})_m.$ N.B. the $\beta$ and $g_\beta$ are part of the specification of the strong curve.  The corresponding weak notion of curve only asks for the existence of $\beta$ and $g_\beta$, but does not specify them

A \emph{(free) path} in $\mathfrak{A}$ will be a curve that stabilises to a constant value on both its left and right ends, i.e. it is an $f : \mathfrak{L} \to \mathfrak{A}$ such that there are integers $N^- \leq N^+$ with the property that \\
\centerline{for all $n \leq N^-$, $f(n) = f(N^-)$,}
\centerline{for all $n \geq N^+$, $f(n) = f(N^+)$,}
Of more use for modelling runs is the notion of a \emph{based path} (i.e. when a basepoint / initial state is specified, but no final point is mentioned).  Given a basepoint $a_0 \in \mathfrak{A}$, a based path in $(\mathfrak{A},a_0)$ is a `free' path that stabilises to $a_0$ on the left, i.e. $f(N^-) = a_0$.  We can similarly define a based curve by requiring merely left stabilisation at $a_0$.  Runs correspond to such curves in which $N^- = 1$.

\section{Objects of Curves, and Paths}
Within the interpreted systems approach to MASs, a set $R$ of runs is often considered as a model (see, for example, \cite{HM:KCKDE} or \cite{HM:UCKBP99}).  The equivalence frame structure given to $R$ may involve the local history of each processor / agent or merely the various `points' visited at the same time; see the discussion in \cite{MvdH} p.39.  The groupoid atlas viewpoint provides a local frame structure on $R$ that is canonical, but, of course, that will need evaluating for its relevance to the problems of MASs.

Let $\mathfrak{A}$ be a groupoid atlas with coordinate system $\Phi_\mathfrak{A}$, underlying set $X_\mathfrak{A}$, etc, as before.  We will write $\textsc{Curves}(\mathfrak{A})$ for the set of curves in $\mathfrak{A}$.

If $f : \mathfrak{L}\to \mathfrak{A}$ is a curve in $\mathfrak{A}$, a function $\beta : |\mathfrak{L} | \to \Phi_\mathfrak{A}$ \emph{frames} $f$ if $\beta$ is a function such that\\
(i) for $m \in |\mathfrak{L}|$, $f(m) \in (X_\mathfrak{A})_{\beta(m)}$;\\
(ii) for $m \in |\mathfrak{L}|$, there is a $b$ in $\Phi_\mathfrak{A}$ with $b\geq \beta(m)$, $b \geq \beta(m+1)$ and a $f(i_m) : f(m) \to f(m+1)$ in $(\mathcal{G}_\mathfrak{A})_b$.

\medskip

\textbf{Remarks:}\\
(a)  The intuition is that the local set containing $f$ in $\textsc{Curves}(\mathfrak{A})$ will consist of curves passing through the same local sets $(X_\mathfrak{A})_\alpha$ in the same sequence.  The idea of a framing of $f$ is that $\beta$ picks out the local sets $(X_\mathfrak{A})_{\beta(m)}$ that receive $f(m)$.  Condition (ii) then ensures that these choices are compatible with the requirement that $f$ be a curve.\\
(b) We have used several times the groupoid atlas associated to a Kripke frame or SGS.  The set $\Phi$ in that case was just the set of agents with the discrete order.  This use of the discrete order is too simplistic in general as it hides the relationships between the agents.  Mathematically this simple model breaks down first on considering framings, since the condition (ii) implies $b\geq \beta(m)$ and $b\geq \beta(m+1)$ so $\beta(m) = \beta(m+1)$ if the order is discrete, but then if $\beta$ is to frame $f$, $f$ must never have left a single equivalence class of the Kripke frame which was not the intention!  To avoid this silly restriction, we can replace the set of agents by the finite non-empty subsets of that set.

\medskip

\textbf{Kripke frames to Groupoid Atlases revisited.}

Suppose  $F = (X, \sim_1, \ldots, \sim_n)$  is a Kripke frame.  Define a new groupoid atlas $\mathfrak{A}^\prime(F)$ by :

$|X_{\mathfrak{A}^\prime(F)}| = X$, the underlying set of $F$;

$\Phi_{\mathfrak{A}^\prime(F)}$ = the set of non-empty subsets of $A$ ordered by $\supseteq$, i.e. $\alpha \leq \beta$ if $\alpha \supseteq \beta$;

$|X_{\mathfrak{A}^\prime(F)}|_\alpha  = X$, for all $\alpha \in \Phi_{\mathfrak{A}^\prime(F)}$\\
and

$\sim_\alpha = \bigcap \{ \sim_i : i \in \alpha\}$, i.e. the equivalence relation
$$x\sim_\alpha y\Leftrightarrow \bigwedge_{i \in \alpha} ( x \sim_i y)$$

\textbf{Remark.}

We can think of $\mathfrak{A}^\prime(F)$ as a `subdivision' of
$\mathfrak{A}(F)$, rather like the barycentric subdivision of a simplicial
complex, a construction to which it is very closely related. To any global
action or groupoid atlas, one can assign two simplicial complexes; see Appendix.  These encode valuable geometric information about the system and relate to the interaction of the different equivalence classes. (Fuller details can be found in \cite{GAGA}.)  Our subdivision above makes no significant change to the homotopy information encoded in the corresponding complexes.

This subdivision is just what is needed to encode runs in `framings'.  Logically, it seems to correspond to the enrichment of our language with `group common knowledge' operators $K_\alpha$, $\alpha\subseteq A$, or dually `group possibility' operators $M_\alpha$, where
$$K_\alpha \phi = \bigwedge_{i\in \alpha}K_i\phi, \mbox { etc.}$$
Here it should be possible to adapt the `subdivision' to reflect more closely the geometry of the distributed system.  For instance, not all finite sets of agents might be included as there might be no direct link between certain of them.  The clique complexes used in analyses of scheduling  problems in distributed systems and in the theory of traces may be relevant here.

\medskip

Now let $\mathfrak{A}$ be a general groupoid atlas and let $\mathfrak{A}^\mathfrak{L}$ be the following data for a groupoid atlas:

$|X_{\mathfrak{A}^\mathfrak{L}}| = \textsc{Curves}(\mathfrak{A})$;

$\Phi_{\mathfrak{A}^\mathfrak{L}} = \{\beta : |\mathfrak{L}| \to \Phi_\mathfrak{A} ~|~ \beta \mbox{  frames some curve }  f \mbox{ in } \mathfrak{A}\}$ \\
For $\beta \in \Phi_{\mathfrak{A}^\mathfrak{L}}$, 

$(X_{\mathfrak{A}^\mathfrak{L}})_\beta = \{f  \in \textsc{Curves}(\mathfrak{A}) ~|~ \beta \textrm{ frames } f\}$

$(\mathcal{G}_{\mathfrak{A}^\mathfrak{L}})_\beta = \{ (\sigma_m) ~|~ \textrm{source}(\sigma_m) \in (X_{\mathfrak{A}^\mathfrak{L}})_\beta, \sigma_m \in (\mathcal{G}_{\mathfrak{A}^\mathfrak{L}})_{\beta(m)}\}$

Note that it is easy to see that $\textrm{target}(\sigma_m) $ is also in $(X_{\mathfrak{A}^\mathfrak{L}})_\beta$ in this situation (see lemma in section 4 of \cite{GAGA}).

Finally define
$$\beta\leq \beta^\prime \Leftrightarrow  \beta(m)\leq \beta^\prime(m) \textrm{ for all } m \in |\mathfrak{L}|.$$

\begin{proposition}~\\
With the above notation,  $\mathfrak{A}^\mathfrak{L}$ is a groupoid atlas. If $\mathfrak{A}$ is a global action, then so is  $\mathfrak{A}^\mathfrak{L}$ .\hfill$\blacksquare$
\end{proposition}
It is natural to ask if $\mathfrak{A} =  \mathfrak{A}^\prime(F)$ for $F$ a Kripke frame, is $\mathfrak{A}^\mathfrak{L}$ associated to some Kripke frame.  In general the answer would seem to be no as there will be more than one local `patch', $(X_{\mathfrak{A}^\mathfrak{L}})_\beta$, in this case and the index set is that of \emph{all} framings.  In fact this structure is not obviously in the MAS literature.  Each framing $\beta$ of a curve $f$ in $\mathfrak{A}^\prime(F)$ defines a sequence $(\beta(m))$ of finite non-empty subsets of the set of agents, satisfying the Kripke frame version of condition (ii) namely that if $m \in |\mathfrak{L}|$, there is a $b$ with $b \geq \beta(m)$, $b \geq \beta(m+1)$ and $f(m) \sim_b f(m+1)$.  In other words $b \subseteq \beta(m) \cap \beta(m+1)$ and $ f(m) \sim_b f(m+1)$.  For a given set of runs, the framings \emph{may} reflect a possiblility of some modularisation as they indicate which agents are idle during the run.  This raises an interesting problem of using the framings to optimise use of resources.

Each of the local groupoids in $\mathfrak{A}^\prime(F)^\mathfrak{L}$ is an equivalence relation on that local patch.  Given $f$, $f^\prime \in (X_{ \mathfrak{A}^\prime(F)})_\beta$, so $\beta$ frames  both $f$ and $f^\prime$, they will be equivalent if $$f(m) \sim_{\beta(m)} f^\prime(m)$$
for each $m$.  These linked pairs together with the fact that $f(m) \sim_bf(m+1)$ and $ f^\prime(m) \sim_{b^\prime}f^\prime(m+1)$, for some $b, b^\prime \subseteq\beta(m) \cap \beta(m+1)$ give a pattern rather like a ladder of linked `squares'.  This is more or less a `homotopy' between $f$ and $f^\prime$ .

\textbf{Remark}

Perhaps a framing can best be thought of as the sequence of those subsets of $A$ involved in a computation at each instant.  Thus a particular agent may be idle throughout a run if no framing of that run/curve involves that agent.  Sometimes more than one agent is involved in a transition at a particular time step, so if, at time $m$, the corresponding set of agents is $\beta(m)$, this interprets as saying that the two global states $f(m) $ and $f(m+1)$ are $\beta(m)$-equivalent, i.e. $f(m)\sim_if(m+1)$ for \emph{all} the agents $i$ in $\beta(m)$.  (In a SGS, which we can imagine as a hypercube for simplicity, then, for example, if $n=5$, $f(m) = (s_1,s_2,s_3,s_4,s_5)$, and  $f(m) = (s_1,s_2,s_3,s_4^\prime,s_5^\prime)$ are $\{1,2,3\}$-equivalent.)
\medskip

The notion of homotopy between based curves should correspond to that of a path in
$\mathfrak{A}^\mathfrak{L}$, where we \emph{do} need `path' not curve so that it stabilises to the given two curves at the two `ends' of that path.  We will not explore this here due to lack of space. The basics of a general treatment of homotopy for groupoid atlases can be found in \cite{GAGA} and in more detail in \cite{EGM} (from the point of view of a cylinder based theory as against a cocylinder theory as would be natural from the viewpoint we have explored here).  Another extremely useful source for this type of theory is \cite{MG1}.

\medskip

\textbf{Cartesian closedness?}

Clearly the object of paths as defined corresponds to a mapping object with domain $\mathfrak{L}$.  This raises the important but difficult question of the cartesian closedness of the category of groupoid atlases and more importantly of the part of it corresponding to the Kripke frames.  Bak has shown \cite{AB1,AB2} that global actions do allow a function space construction that is well behaved on a large class of examples.  A closely related construction occurs with equilogical spaces as defined by Scott, \cite{Scott}.  These are $T_0$-spaces together with an equivalence relation.  Kripke equivalence frames for a `single agent system'  give equilogical spaces and equilogical spaces form a cartesian closed category.  No analogues of equilogical spaces for systems of $n$-agents seem to have been developed. Similarly no analogues are known where different models have different numbers of `agents'. Yet from a logical point of view and for an an adequate logical language to handle multiagent systems, some setting in which a cartesian closed category structure is available is clearly desirable.

\section{Conclusions, Critique and Future Directions}

In this paper, I have tried to examine some of  the methodological links between the theory of global actions / groupoid atlases and the general context of combinatorial models for studying multiagent systems.  Within the space available, no firm conclusions can be reached as to the potential usefulness of these links, but the possibility of a better structured object of runs in a distributed system has been shown that extends the Kripke frames of runs considered in the MAS literature.

What has not been done?  It is clear that a more detailed examination of homotopy is required, especially with respect to its interpretation in terms of computation.  The problem of cartesian closedness has been noted, but, deliberately, set aside due to lack of space  and firm knowledge.  The whole question of the relationship between these constructions and bounded morphisms (and thus with the logic) has also been set aside. It is conjectured that bounded morphisms will form a (subclass of the) class of `fibrations' in the homotopy theory of this context, since, for a bounded morphism $f : A \to B$ of frames, for each $i$, the condition corresponds to being a fibration of groupoids.

Finally, but crucially, the computational infeasibility of $S5_n$ suggests that a separate study using $S4_n$, probably in an intuitionistic form, cf. \cite{EG&JGL1,EG&JGL2}, will be worth doing.  This will presumably need a version of directed homotopy, but which variant of the many available, cf. \cite{EG1}, will best suit is not yet clear.  (The ideas in \cite{MG1} are also very relevant here.) Perhaps then, some deeper evaluation of how the geometry of a distributed system influences its inherent logic and thus its computational ability will become possible.

\pagebreak

\appendix

\section*{Appendix : Simplicial complexes from Kripke frames}

The two constructions mentioned in the main text are classical, dating in their initial forms to the
embryonic algebraic topology of the 1920s and 30s.  The local
equivalence classes of a Kripke frame give a covering of the underlying set, $X$ of the frame.  It thus gives a relation from $X$ to the set $Y$ of
equivalence classes.  Abstracting, let $R\subset X \times Y$ be a relation. (In
our case $xRy$ is exactly $x\in y$, where $y$ is an equivalence class for any
of the equivalence relations).  Using a formulation due to Dowker,
\cite{dowker}, any such relation determines two simplicial complexes

\begin{enumerate}
\item[(i)] $K = K_R :$
\begin{enumerate}
\item[-]the set of vertices is the set, $Y$;
\item[-]  $p$-simplex of $K$ is a set $\{y_0, \cdots, y_p\}
  \subseteq Y$ such that there is some $x \in X$ with $x Ry_j$ for $j
  = 0, 1,~ \cdots,~ p$.
\end{enumerate}
\item[(ii)] $L = L_R:$
\begin{enumerate}
\item[-]the set of vertices is the set $X$;
\item[-] a $p$-simplex of $K$ is a set $\{x_0, \cdots, x_p\}
  \subseteq X$ such that there is some $y \in Y$ with $x_i Ry$ for $i
  = 0, 1,~ \cdots, ~p$.
\end{enumerate}
\end{enumerate}
These are, in some sense, dual constructions.  In the topological context,
$K_R$ is often called the \emph{Nerve} of the covering and $L_R$ the
\emph{Vietoris  complex}.  

As a simple example, let $X = \{1,2, \ldots ,6\}$,
\\
$a\sim_1 b$ if $a - b$ is a multiple of 2;\\
$a\sim_2 b$ if $a - b$ is a multiple of 3.

This corresponds to a hypercube, $L_1\times L_2$, with $L_1$ having 3 elements, $L_2$ having
2.  $Y$ has 5 elements. $X$ has 6.

$K_R$ is a bipartite graph: 
\vspace{.5cm}

$$
\xymatrix{
\{1,4\} \ar@{-}[rd] \ar@{-}[rrrd]&& \{3,6\}\ar@{-}[ld]\ar@{-}[rd] && \{2,5\} \ar@{-}[llld]
\ar@{-}[ld] \\
&\{1,3,5\}&& \{2,4,6\} &  
}
$$
\vspace{.5cm}

$L_R$ is a prism with two filled triangular faces:

\begin{figure}[h]
\begin{center}
\includegraphics[scale = 0.7]{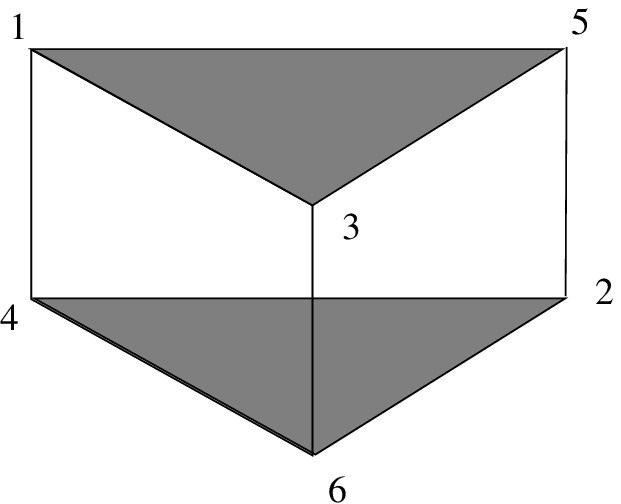}
\end{center}
\end{figure} 

They have both the homotopy type of a figure 8. (For instance in the prism, shrink the triangles to points and then shrink  one vertical edge.) 

The main result of Dowker's paper was that for an arbitrary relation $R$, the two complexes have the same homotopy type.  The question of the influence of the homotopy type of these complexes on the complexity of searches in the state space of the original MAS seems to be a very interesting one.
\end{document}